\documentclass{article}
\usepackage{graphicx}
\usepackage{cite}

\begin{document}

\title{On the crossing of the energy levels of a parameter-dependent
quantum-mechanical Hamiltonian}
\author{Francisco M. Fern\'{a}ndez\thanks{%
E-mail: fernande@quimica.unlp.edu.ar} \\
INIFTA (UNLP, CCT La Plata-CONICET), Divisi\'{o}n Qu\'{i}mica Te\'{o}rica \\
Blvd. 113 S/N, Sucursal 4, Casilla de Correo 16 \\
1900 La Plata, Argentina}
\date{}
\maketitle

\begin{abstract}
The non-crossing rule for the energy levels of a parameter-dependent
Hamiltonian is revisited and a flaw in a commonly accepted proof is
revealed. Some aspects of avoided crossings are illustrated by means of
simple models. One of them shows the close relationship between avoided
crossings and exceptional points.
\end{abstract}

\section{Introduction}

It is well known that the potential energy curves of diatomic molecules do
not cross (in fact, they even avoid each other) when the states have the
same symmetry. This property of the electronic energies, commonly known as
the non-crossing rule, has proved useful for the interpretation of many
experiments in molecular spectroscopy and photochemistry\cite{D74,H50}. The
theoretical explanation outlined by Teller\cite{T37}, and typically
reproduced in most textbooks on quantum chemistry\cite{EWK44, P68}, was
criticized by Razi Naqvi and Byers Brown\cite{RNBB72}. After arguing that
such a proof is based on a \textit{non sequitur} the authors proposed an
alternative justification of the non-crossing rule. Their argument is
closely related to the Hellmann-Feynman theorem\cite{F39} in its more
general off-diagonal form\cite{KD78, CF80, F04, BHM04, V04}.

As a result of a recent investigation on non-Hermitian Hamiltonians we were
led to revise the proofs on the non-crossing rule and the purpose of this
paper is to put forward our analysis and discussion of the arguments given
by Razi Naqvi and Byers Brown\cite{RNBB72}. In section~\ref{sec:H-F} we
derive similar equations by means of the off-diagonal Hellmann-Feynman
theorem\cite{KD78, CF80, F04, BHM04, V04} (and references therein). In
section~\ref{sec:Examples} we illustrate the main theoretical conclusions by
means of two simple examples. Finally, in section~\ref{sec:Conclusions} we
summarize the main results and draw conclusions.

\section{The off-diagonal Hellmann-Feynman theorem}

\label{sec:H-F}

The starting point is the time-independent Schr\"{o}dinger equation
\begin{equation}
H\psi _{n}=E_{n}\psi _{n}.  \label{eq:Schro}
\end{equation}
It follows from $\left\langle \psi _{m}\right| H\left| \psi
_{n}\right\rangle =\left\langle \psi _{n}\right| H\left| \psi
_{m}\right\rangle ^{*}$ and $E_{m}^{*}=E_{m}$, where * stands for complex
conjugation, that
\begin{equation}
\left( E_{m}-E_{n}\right) \left\langle \psi _{m}\right. \left| \psi
_{n}\right\rangle =0.  \label{eq:pre_orthogonality}
\end{equation}
From this expression we conclude that $\left\langle \psi _{m}\right. \left|
\psi _{n}\right\rangle =0$ when $E_{m}\neq E_{n}$. This textbook result is
well known but we write it here because it will be useful later on.

If $H$ depends on a parameter $\lambda $, then the eigenfunctions and
eigenvalues will depend on this parameter too. If we differentiate equation (%
\ref{eq:Schro}) with respect to $\lambda $ and then apply the bra $%
\left\langle \psi _{m}\right| $ from the left we obtain the well known
off-diagonal Hellmann-Feynman relation\cite{KD78, CF80, F04, BHM04, V04}
\begin{equation}
\left\langle \psi _{m}\right| H^{\prime }\left| \psi _{n}\right\rangle
=E_{n}^{\prime }\left\langle \psi _{m}\right. \left| \psi _{n}\right\rangle
+\left( E_{n}-E_{m}\right) \left\langle \psi _{m}\right. \left| \psi
_{n}^{\prime }\right\rangle ,  \label{eq:H-F}
\end{equation}
where the prime denotes differentiation with respect to $\lambda $.

Suppose that $E_{m}$ and $E_{n}$ approach each other and cross at $\lambda
_{0}$: $\lim\limits_{\lambda \rightarrow \lambda _{0}}\left(
E_{m}-E_{n}\right) =0$. When $\lambda \neq \lambda _{0}$ $\left\langle \psi
_{m}\right. \left| \psi _{n}\right\rangle =0$ by virtue of equation (\ref
{eq:pre_orthogonality}) and because of continuity we should also have
\begin{equation}
\lim\limits_{\lambda \rightarrow \lambda _{0}}\left\langle \psi _{m}\right.
\left| \psi _{n}\right\rangle =0.  \label{eq:orthogonality}
\end{equation}
It follows from this equation and (\ref{eq:H-F}) that
\begin{equation}
\left\langle \psi _{m}\right| H^{\prime }\left| \psi _{n}\right\rangle
(\lambda _{0})=\lim\limits_{\lambda \rightarrow \lambda _{0}}\left\langle
\psi _{m}\right| H^{\prime }\left| \psi _{n}\right\rangle =0.
\label{eq:H'mn=0}
\end{equation}
Without this condition the approaching energy levels will not cross giving
rise to an avoided crossing that looks like an energy-level repulsion. Since
the two levels approach each other and then move apart the quantity $\left(
E_{n}-E_{m}\right) ^{2}$ should exhibit a minimum at some $\lambda =\lambda
_{m}$. This particular value of the parameter is determined by the condition
\begin{equation}
E_{n}^{\prime }(\lambda _{m})-E_{n}^{\prime }(\lambda _{m})=0
\label{eq:E'_n-E'_m=0}
\end{equation}
If the symmetries of $\psi _{m}$ and $\psi _{n}$ are different, then
equation (\ref{eq:H'mn=0}) holds for all $\lambda $ and nothing prevents the
approaching energy levels from crossing.

Throughout the discussion above we have tacitly assumed that the symmetry of
$H$ is the same for all $\lambda $ (at least in the neighbourhood of $%
\lambda _{0}$ under analysis). In other words, we have assumed that both $H$
and $H^{\prime }$ have the same symmetry. Suppose that the point group\cite
{C90,T64} that describes the symmetry of $H$ is $G$ when $\lambda \neq
\lambda _{0}$ and $G_{0}$ when $\lambda =\lambda _{0}$ and that the order $h$
of $G$ is smaller than the order $h_{0}$ of $G_{0}$. Under such conditions
the dimension of the subspaces of $H(\lambda _{0})$ may be greater than
those for $H($ $\lambda \neq \lambda _{0})$ and we therefore expect some
level crossings at $\lambda =\lambda _{0}$. Obviously, equation (\ref
{eq:H'mn=0}) applies to those states that become degenerate at this point.
Razi Naqvi\cite{RN72} took into account such symmetry changes in a
discussion of the crossing of potential-energy surfaces of polyatomic
molecules.

It is clear that no further discussion is necessary for proving equation (%
\ref{eq:orthogonality}) that was required for deriving equation (\ref
{eq:H'mn=0}) from (\ref{eq:H-F}). However, Razi Naqvi and Byers Brown\cite
{RNBB72} criticized the continuity argument implied by equation (\ref
{eq:orthogonality}). In order to discuss the additional steps in their proof
we first derive another equation. If we differentiate the eigenvalue
equation for $\psi _{m}$ with respect to $\lambda $ and apply $\left\langle
\psi _{n}\right| $ from the left we arrive at an equation similar to (\ref
{eq:H-F}):
\begin{equation}
\left\langle \psi _{n}\right| H^{\prime }\left| \psi _{m}\right\rangle
=E_{m}^{\prime }\left\langle \psi _{n}\right. \left| \psi _{m}\right\rangle
+\left( E_{m}-E_{n}\right) \left\langle \psi _{n}\right. \left| \psi
_{m}^{\prime }\right\rangle .  \label{eq:H-F_2}
\end{equation}
Subtracting the complex conjugate of equation (\ref{eq:H-F_2}) from equation
(\ref{eq:H-F}) we obtain
\begin{equation}
\left( E_{n}-E_{m}\right) ^{\prime }\left\langle \psi _{m}\right. \left|
\psi _{n}\right\rangle +\left( E_{n}-E_{m}\right) \left\langle \psi
_{m}\right. \left| \psi _{n}\right\rangle ^{\prime }=0,
\label{eq:deriv_orthog}
\end{equation}
which is obviously the derivative of equation (\ref{eq:pre_orthogonality})
with respect to $\lambda $. When $\lim\limits_{\lambda \rightarrow \lambda
_{0}}\left( E_{m}-E_{n}\right) =0$ equation (\ref{eq:deriv_orthog}) reduces
to
\begin{equation}
\left[ E_{n}^{\prime }(\lambda _{0})-E_{m}^{\prime }(\lambda _{0})\right]
\left\langle \psi _{m}\right. \left| \psi _{n}\right\rangle (\lambda _{0})=0.
\label{eq:RNBB1}
\end{equation}

Razi Naqvi and Byers Brown\cite{RNBB72} considered two electronic states $%
\psi _{1}$ and $\psi _{2}$ of a diatomic molecule such that the
corresponding electronic energy levels $E_{1}(R)$ and $E_{2}(R)$, where $R$
is the internuclear distance, cross at $R=R_{0}$. They derived an equation
similar to (\ref{eq:RNBB1}) that reads:
\begin{equation}
\left[ E_{1}^{\prime }(R_{0})-E_{2}^{\prime }(R_{0})\right] \left\langle
\psi _{1}^{0}\right. \left| \psi _{2}^{0}\right\rangle =0,  \label{eq:RNBB2}
\end{equation}
where $\psi _{1}^{0}$ and $\psi _{2}^{0}$ are the electronic states at $%
R=R_{0}$. They invoked this equation to prove that
\begin{equation}
\left\langle \psi _{1}^{0}\right. \left| \psi _{2}^{0}\right\rangle =0
\label{eq:ortho_RNBB}
\end{equation}
if $E_{1}^{\prime }(R_{0})\neq E_{2}^{\prime }(R_{0})$. The reason of this
\textit{circumlocution} was their concern about the continuity argument
expressed in the statement: ``It will be well to pause here momentarily and
discuss the implications of Equation (\ref{eq:ortho_RNBB}). Our demand that
the two potential curves intersect at $R_{0}$, forces us, to conclude that
the overlap integral must vanish even when $E_{1}=E_{2}$. It is tempting to
argue that, since $\left\langle \psi _{1}\right. \left| \psi
_{2}\right\rangle =0$ for all $R$ in the vicinity of $R=R_{0}$, it seems
likely, on account of continuity, that it would also be true at $R_{0}$.
However, this argument is not only unnecessary but misleading, for we know
that degenerate eigenfunctions need not be orthogonal; indeed we can choose
them at will and make them to be non-orthogonal, if we so desire.'' The
reader may convince himself that the argument leading to equation (\ref
{eq:orthogonality}) clearly implies that we do not choose those functions
``at will'' because the states at $R=R_{0}$ are just the ones that result
from the limit $R\rightarrow R_{0}$ and, therefore, should remain orthogonal.

In order to prove that equation (\ref{eq:ortho_RNBB}) holds even when $%
E_{1}^{(j)}(R_{0})=E_{2}^{(j)}(R_{0})$, $j=0,1,\ldots ,n$, provided that $%
E_{1}^{(n+1)}(R_{0})\neq E_{2}^{(n+1)}(R_{0})$, the authors differentiate
equation (\ref{eq:RNBB2}) with respect to $R$ as many times as necessary\cite
{RNBB72}. However, it is obvious that equation (\ref{eq:RNBB2}) is valid
only for $R=R_{0}$ because we have discarded a term from the general
equation valid for all $R$ (see (\ref{eq:deriv_orthog})). In order to carry
out this proof correctly we should differentiate an equation like (\ref
{eq:deriv_orthog}) as many times as necessary which is equivalent to
differentiating an equation similar to (\ref{eq:pre_orthogonality}) with
respect to $R$ just one more time. More precisely, if we define $\Delta
(R)=E_{1}(R)-E_{2}(R)$ and $S(R)=\left\langle \psi _{1}\right. \left| \psi
_{2}\right\rangle (R)$ then equation (\ref{eq:pre_orthogonality}) becomes $%
\Delta (R)S(R)=0$. Differentiating it $n+1$ times with respect to $R$ and
substituting $R_{0}$ for $R$ we obtain
\begin{equation}
\sum_{j=0}^{n+1}\Delta ^{(j)}(R_{0})S^{(n+1-j)}(R_{0})=\Delta
^{(n+1)}(R_{0})S(R_{0})=0,  \label{eq:add_proof_ort}
\end{equation}
from which it follows that $S(R_{0})=0$ when $\Delta ^{(n+1)}(R_{0})\neq 0$.
In addition to being simpler and clearer, this argument is free from the
flaw in the additional steps of the proof attempted by Razi Naqvi and Byers
Brown\cite{RNBB72}. However, in our opinion this discussion is unnecessary
because, as argued above, $\lim\limits_{R\rightarrow R_{0}}S(R)=0$ always
applies when $\lim\limits_{R\rightarrow R_{0}}\Delta (R)=0$.

\section{Examples}

\label{sec:Examples}

In section~\ref{sec:H-F} we mentioned the possibility that the symmetry of
the system may change at $\lambda =\lambda _{0}$. In order to illustrate
this point here we choose an extremely simple model, the quantum-mechanical
harmonic oscillator
\begin{equation}
H=-\frac{\partial ^{2}}{\partial x^{2}}-\frac{\partial ^{2}}{\partial y^{2}}%
+kx^{2}+\lambda y^{2},\;k,\lambda >0.  \label{eq:HO_2D}
\end{equation}
The eigenvalues and eigenfunctions of this dimensionless Hamiltonian
operator are given by
\begin{eqnarray}
E_{mn} &=&\sqrt{k}(2m+1)+\sqrt{\lambda }(2n+1),\;m,n=0,1,\ldots ,  \nonumber
\\
\psi _{mn}(x,y) &=&\phi _{m}(x,k)\phi _{n}(y,\lambda ),  \label{eq:HO_eigens}
\end{eqnarray}
respectively, where $\phi _{m}(q,k)$ is an eigenfunction of the
one-dimensional harmonic oscillator $H_{HO}=-\frac{\partial ^{2}}{\partial
q^{2}}+$ $kq^{2}$.

When $\lambda \neq \lambda _{0}=k$ the symmetry of the system is described
by the Abelian point group $C_{2v}$ that exhibits only one-dimensional
irreducible representations\cite{C90,T64}. Therefore, its states are
expected to be nondegenerate, except for accidental degeneracies that may
occur when $\sqrt{\lambda /k}$ is rational. On the other hand, when $\lambda
=\lambda _{0}$ we have an isotropic two-dimensional oscillator so that its
symmetry is described by the full two-dimensional rotation group. Since all
its eigenstates $\psi _{m+j\,n-j}$, $j=-m,-m+1,\ldots ,n$ are degenerate we
expect and infinite number of crossings at $\lambda =\lambda _{0}$.
Obviously, the off-diagonal matrix elements
\begin{equation}
\left\langle \psi _{mn}\right| H^{\prime }\left| \psi
_{m+j\,n-j}\right\rangle =\left\langle \phi _{n}\right| y^{2}\left| \phi
_{n-j}\right\rangle \delta _{m\,m+j},
\end{equation}
vanish when $\lambda =\lambda _{0}$ in agreement with the argument given is
section \ref{sec:H-F}.

\begin{figure}[]
\begin{center}
\includegraphics[width=6cm]{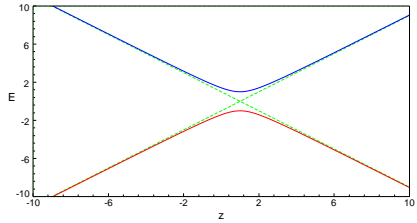}
\end{center}
\caption{Diagonal elements (dashed, green line) and eigenvalues (red, blue,
continuous lines) for the two-level model ( \ref{eq:H_Hermit})}
\label{fig:EH}
\end{figure}

\begin{figure}[]
\begin{center}
\includegraphics[width=6cm]{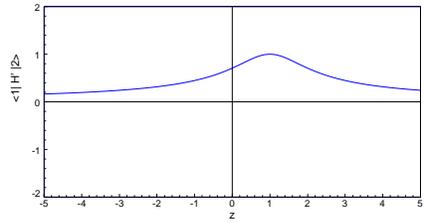}
\end{center}
\caption{Matrix element $\langle \psi_1 | H^{\prime} | \psi_2 \rangle$ for
the Hamiltonian (\ref{eq:H_Hermit})}
\label{fig:H'12}
\end{figure}

The second example is even simpler but most interesting in some respects. In
this case we choose a two-level system given by the matrix representation
\begin{equation}
\mathbf{H}=\left(
\begin{array}{ll}
-1+z & -1 \\
-1 & 1-z
\end{array}
\right) .  \label{eq:H_Hermit}
\end{equation}
The diagonal elements intersect at $z=1$ but the eigenvalues
\begin{equation}
E_{1}=-\sqrt{z^{2}-2z+2},\;E_{2}=\sqrt{z^{2}-2z+2},  \label{eq:E1,E2_Hermit}
\end{equation}
exhibit an avoided crossing as shown in Figure~\ref{fig:EH}. Figure~\ref
{fig:H'12} shows that the off-diagonal matrix element $\mathbf{\psi }%
_{1}^{t}.\mathbf{H}^{\prime }.\mathbf{\psi }_{2}$, where $\mathbf{\psi }_{1}$
and $\mathbf{\psi }_{2}$ are the two column eigenvectors of $\mathbf{H}$,
non only does not vanish but even exhibits a maximum precisely at $z=1$.

\begin{figure}[]
\begin{center}
\includegraphics[width=6cm]{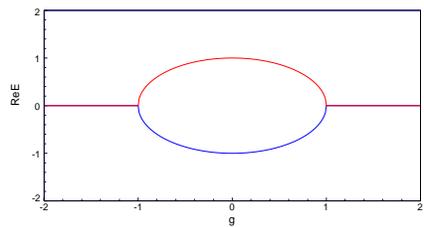}
\end{center}
\caption{Real parts of the two eigenvalues of the two-level model (\ref
{eq:K_non-Hermit})}
\label{fig:ENH}
\end{figure}

It is well known that avoided crossings are associated to exceptional points
in the complex plane\cite{HS90, H00, HH01, H04}. Present case is not an
exception as the eigenvalues (\ref{eq:E1,E2_Hermit}) obviously cross in the
complex $z$-plane at $z=1\pm i$. By means of the change of variables $z=1+ig$
we obtain a parity-time-symmetric non-Hermitian Hamiltonian\cite{B07}:
\begin{equation}
\mathbf{K}=\left(
\begin{array}{ll}
ig & -1 \\
-1 & -ig
\end{array}
\right) ,  \label{eq:K_non-Hermit}
\end{equation}
with eigenvalues
\begin{equation}
E_{1}=-\sqrt{1-g^{2}},\;E_{2}=\sqrt{1-g^{2}}.  \label{eq:E1,E2_non-Hermit}
\end{equation}
In this case the eigenvalues are real for all $|g|<1$ (unbroken parity-time
(PT) symmetry\cite{B07}) approach each other as $g\rightarrow \pm 1$,
coalesce at the exceptional points $g=\pm 1$ and become a pair of complex
conjugate numbers for $|g|>1$ (broken PT symmetry). This behaviour is shown
in Figure~\ref{fig:ENH}. At the exceptional points the two eigenvectors are
linearly dependent\cite{HS90, H00, HH01, H04}.

\begin{figure}[]
\begin{center}
\includegraphics[width=6cm]{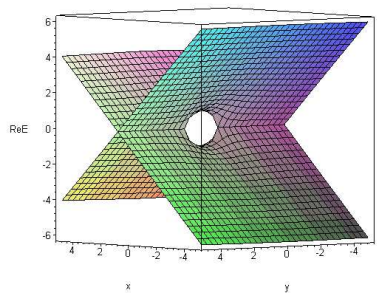}
\par
\end{center}
\caption{Real parts of the eigenvalues of the Hamiltonian (\ref{eq:H_Hermit}%
) when $z=x+iy$}
\label{fig:3D}
\end{figure}

It is most interesting to consider the more general case in which $z=x+iy$
that leads to an Hermitian Hamiltonian when $y=0$ and a PT-symmetric one
when $x=1$. Figure~\ref{fig:3D} shows that $\Re E(x,y)$ is given by two
intersecting surfaces that leave a hole where they do not touch. The
intersection of the whole composite surface with the plane $(x,0,z)$ yields
the curves shown in figure~\ref{fig:EH} for the Hermitian Hamiltonian (\ref
{eq:H_Hermit}). On the other hand, the intersection with the plane $(1,y,z)$
yields the curve in figure~\ref{fig:ENH} for the eigenvalues of the
PT-symmetric Hamiltonian (\ref{eq:K_non-Hermit}), where $y=g$.

\section{Conclusions}

\label{sec:Conclusions}

The arguments put forward by Razi Naqvi and Byers Brown\cite{RNBB72} are
basically correct, except for the discussion of the orthogonality of the
states at the crossing point $R=R_{0}$. In the first place, there is no
problem with the orthogonality of the states at this point if one chooses
them to be the result of the limit $R\rightarrow R_{0}$. Such states are not
at all arbitrary and conserve their orthogonality even at the point of
degeneracy. If one had any doubt about the orthogonality of the states at
the crossing point one could in fact prove it as shown in equation (\ref
{eq:add_proof_ort}) that is an improvement on the argument given by those
authors that leads to the correct answer but is based on an inadequate
equation.

\section*{Acknowledgements}

This report has been financially supported by PIP No. 11420110100062
(Consejo Nacional de Investigaciones Cientificas y Tecnicas, Rep\'{u}blica
Argentina)

\end{document}